\documentclass[prl,aps,amsmath,amssymb,floatfix,showpacs,reprint,superscriptaddress]{revtex4-1}
\usepackage{graphicx}

\begin{document}

\newcommand{\PHI}[0]{\ensuremath{\boldsymbol{\phi}}}
\newcommand{\dx}[0]{\ensuremath{\Delta{x}}}
\newcommand{\ket}[1]{\ensuremath{|{#1}\rangle}}
\newcommand{\gat}[0]{\ensuremath{g_{\mathrm{at}}}}
\newcommand{\geff}[0]{\ensuremath{g_{\mathrm{eff}}}}
\newcommand{\omegaat}[0]{\ensuremath{\omega_{\mathrm{at}}}}
\newcommand{\omegaeff}[0]{\ensuremath{\omega_{\mathrm{eff}}}}

\title{Nonequilibrium and nonperturbative dynamics of ultrastrong coupling in open lines}

\author{B. Peropadre}
\address{Instituto de F{\'\i}sica Fundamental IFF-CSIC, Calle Serrano
  113b, Madrid E-28006, Spain}

\author{D. Zueco}
\address{Instituto de Ciencia de Materiales de Arag\'on y
  Departamento de F\'{\i}sica de la Materia Condensada,
  CSIC-Universidad de Zaragoza, Zaragoza E-50009, Spain.}
\address{Fundaci\'on ARAID, Paseo Mar\'{\i}a Agust\'{\i}n 36,
  Zaragoza E-50004, Spain}

\author{D. Porras}
\address{Fac. CC. F{\'\i}sicas, Univ. Complutense de Madrid, Madrid E-28040, Spain}

\author{J.~J. Garc{\'\i}a-Ripoll}
\email{jj.garcia.ripoll@csic.es}
\address{Instituto de F{\'\i}sica Fundamental IFF-CSIC, Calle Serrano
  113b, Madrid E-28006, Spain}

\begin{abstract}
 We study the time and space resolved dynamics of a qubit with an Ohmic coupling to propagating 1D photons, from weak coupling to the ultrastrong coupling regime. A nonperturbative study based  on Matrix Product States (MPS) shows the following results: (i) The ground state of the combined systems contains excitations of both the qubit and the surrounding bosonic field. (ii) An initially excited qubit equilibrates through spontaneous emission to a state, which under certain conditions, is locally close to that ground state, both in the qubit and the field. (iii) The resonances of the combined qubit-photon system match those of the spontaneous emission process and also the predictions of the adiabatic renormalization [A.\ J. Leggett \textit{et al}, Rev. Mod. Phys. 59, 1, (1987)]. Finally, a non-perturbative ab-initio calculations show that this physics can be studied using a flux qubit galvanically coupled to a superconducting transmission line.
\end{abstract}

\pacs{Unknown}

\maketitle

Recently achieved in experiments with superconducting circuits\ \cite{Niemczyk2010,Forn-Diaz2010}, polaritons~\cite{Gunter2009,Anappara2009} and two-dimensional electron gases\ \cite{Geiser2012}, Utra-Strong Coupling (USC) is usually linked to the study of discrete systems interacting with cavities, where it is defined as \textit{the coupling strength at which counterrotating terms become relevant}, the number of excitations (photons) is not conserved and the Rotating Wave Approximation (RWA) breaks down.

We will extend the notion of USC to free space, describing superconducting qubits in open transmission lines. For that we model the atom-light interaction with the spin-boson (SB) Hamiltonian\ \cite{Leggett1987}
\begin{equation}
  H = \sum_k \omega_k a^\dagger_k a_k + \frac{\omegaat}{2}\sigma^z + g \sigma^x
  \sum_k (u_k^*  a_k + u_k a_k^\dagger).
  \label{H}
\end{equation}
This contains a quasi-continuum of frequencies for the propagating photons, $\omega_k$, a two-level system for the superconducting qubit, $\omegaat$, and a realistic set of coupling strengths $u_k$ for the specific qubit type. As in the interrupted transmission line\ \cite{LeHur2012}, the qubit-line system belongs to the Ohmic regime, with a linear spectral function
\begin{equation}
  J(\omega) = \pi \sum_k 2g^2 |u_k|^2 \delta(\omega - \omega_k) \sim
  2\pi\alpha \omega^1.\label{spectral-function}
\end{equation}
The parameter $\alpha= 2(\geff/\omegaat)^2$ quantifies the strength of the SB coupling, but it is also related to the coupling $\geff$ between the qubit and a resonant cavity made from the same transmission line. While USC effects will be shown in this work for $\alpha\gtrsim 0.1$ or $\geff/\omegaat \gtrsim 25\%$, a drastic change in the dynamics is observed at the point $\alpha = 1/2$ ($\geff/\omegaat=50\%$) at which \textit{both} non-Markovian and non-RWA effects become relevant in free space.

The goal of this work is precisely to develop theoretical tools for studying the relaxation and scattering dynamics of a superconducting qubit in an open line, in all coupling regimes ---weak, USC and beyond. A proper description of such ongoing and future experiments\ \cite{Astafiev2010,Abdumalikov2010,Hoi2011,Hoi2012} demands theoretical tools that study simultaneously the dynamics of the qubits and the bosons, both in time and space, accurately ---i.e. without tracing out the line or applying ad-hoc decoupling schemes---. This goal is achieved using customized MPS numerical methods that merge ideas from the quantum impurity ansatz\ \cite{Weichselbaum2009}, Matrix Product Operators\ \cite{Murg2008} and mixed time evolution methods\ \cite{Garcia-Ripoll2006}. Our methods rely on the coupled resonator model for the 1D transmission line. Thus, unlike logarithmic discretizations in energy space NRG\ \cite{Anders2005,Anders2006,Orth2010} or polynomial discretization MPS\ \cite{Prior2010}, we strive for precise representations of real space observables, such as the distribution of photons, propagating wave packets or correlations.

A summary of the main results is as follows. We start by computing the vacuum photon fluctuations that arise from the squeezing in the lumped element model, together with the nonlocal distribution of photons that arises when we ultrastrongly couple the qubit to the transmission line. This distribution of photons is different for charge and flux qubits, the later being more delocalized. Next, we study the dynamics of a qubit which is initially excited and which is suddenly coupled to the transmission line. The MPS simulations reveal how the qubit relaxes towards the joint qubit-line ground state by spontaneously emitting a photon that travels away, leaving part of the system in a quasi-stationary state. The qubit-line system, though a closed system, seems to provide a bath for its own equilibration, where the bath are the far away extremes of the line. Nevertheless, in frequency space this equilibration is not obvious, due to the presence of the spontaneously emitted photon, whose frequency shows excellent agreement with adiabatic renormalization theory~\cite{Leggett1987}. The converse experiment, that is the interaction of the qubit with incoming photons shows the same resonance, which could be identified from scattering experiments\ \cite{Astafiev2010,Abdumalikov2010,Hoi2011,Hoi2012} and correlation functions\ \cite{Hoi2012}. Finally, we develop an ab-initio theoretical model for the coupling strength between a flux qubit interacting and an open transmission line. Unlike Ref.~\cite{Bourassa2009} our model is non-perturbative proves that, using parameters from ordinary transmission lines and three-junction qubits, it is possible to achieve all coupling regimes of the SB model, making this a suitable platform to test our predictions.

\begin{figure}[t]
  \centering
  \includegraphics[width=0.8\linewidth]{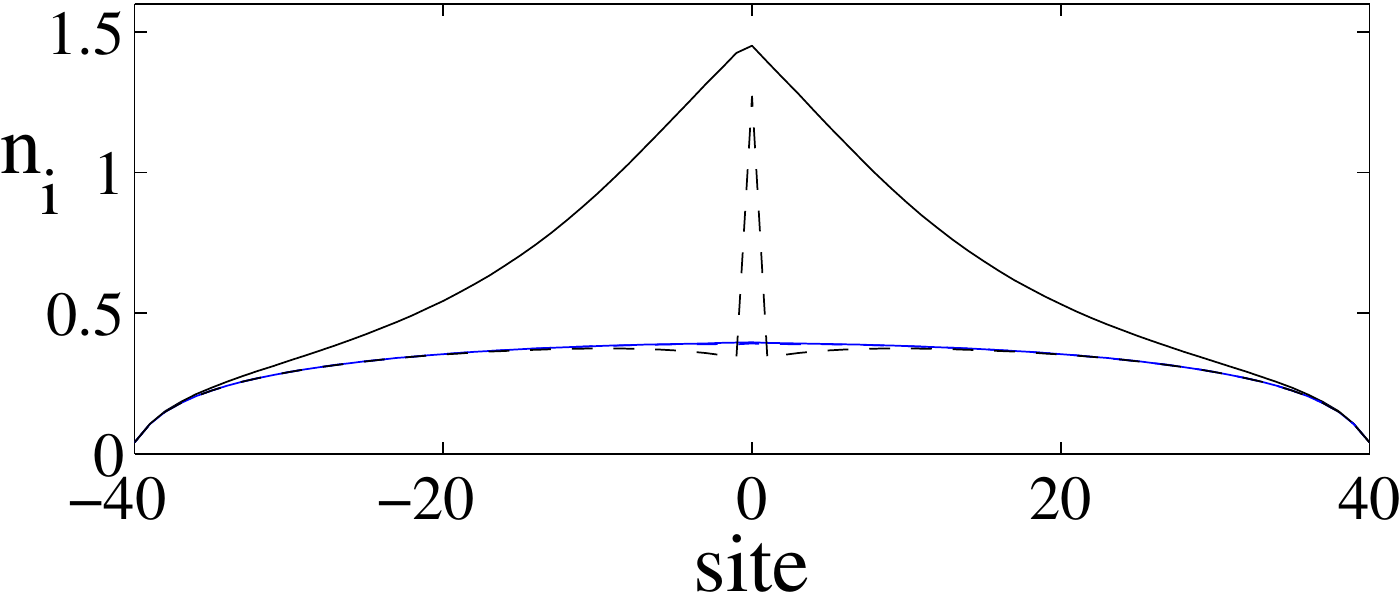}
  \caption{Ground state occupation numbers for a line without a qubit ($\alpha=0$, blue line below), with a flux qubit and couplings of type, $O_x$ ($\alpha=1$, solid peaked line), and charge qubit $O_p$ ($\alpha=1$, dashed). Note how the perturbation of line due to the flux qubit is always extended all over the line.}
  \label{fig:ground-state}
\end{figure}

We write the interaction between a two-level system and a one-dimensional waveguide of photons as
\begin{equation}
  H = \sum_i \frac{\omega_0}{2} [(x_{i+1}-x_i)^2 + p_i^2] + \frac{\omegaat}{2} \sigma^z
  + g\sigma^x O_{p,x}.
  \label{H-position}
\end{equation}
In a superconducting circuit, $x_i$ and $p_i$ are the flux and charge variables, respectively; the set of coupled oscillators is the equivalent circuit for a transmission line\ \cite{Denker1984,Devoret1995} and the coupling will be $O_p=p_0$ or $O_x =x_1-x_0$, for charge and flux qubits, respectively\ \footnote{All simulations have been done using both couplings, without significant differences except where explicitly noted.}. Since the ground state squeezing prevents an efficient MPS description in real space\ \footnote{The entanglement generated by the coupling leads to a large number of photons $n_i=\tfrac{1}{2}(x_i^2+p_i^2-1)$ that demand a large cut-off to represent the state faithfully, even without a qubit.}, we work in frequency space (\ref{H}) using the open boundary conditions modes of a chain of length $L$, with quasimomentum $k=\frac{\pi}{L+1} \times \{1\ldots L\}$ and spectrum $\omega_k = \omega_0 \sqrt{2-2\cos(k)}$. Note that we still may recover expectation values of the $x_i$ and $p_i$ operators using the new Fock operators, $a_k$, and an orthogonal change of basis. Also note that the SB model is characterized by the ultraviolet cut-off $\omega_c = \sqrt{2}\omega_0$ and a parameter $\alpha$\ (\ref{spectral-function}) that can be inferred from a linear fit to the spectrum.

We write any physical state in the MPS form as
\begin{equation}
  \ket{\psi} = \mathrm{tr} (A_0^s A_1^{n_1} \cdots A_L^{n_1})
  \ket{s,n_1,\ldots,n_L},
\end{equation}
where $s=0,1$ is the qubit state and $n_i\in0,1,\ldots,n_{max}$ is the photon occupation number of the different modes. Splitting the Hamiltonian as $H=H_0 + g H_I$, time evolution is approximated using a Trotter formula
\begin{equation}
  \ket{\psi(t)} = \left(e^{-iH_0t/2N} e^{-igH_I t/N} e^{-igH_0t/2N}\right)^N \ket{\psi(0)},
\end{equation}
with a sufficiently small time step $\Delta t = t/N$. This decomposition has an associated MPS algorithm where evolution with $H_0$ is exact and evolution with $H_I$ is approximated using an Arnoldi method~\cite{Garcia-Ripoll2006}, with small truncation errors to preserve the MPS form with a small bond dimension, $\chi=\max_{s,i}\dim A_i^s \sim 10-40$.

%%%%%%%%%%%%%%%%%%%%%%%%%%%%%%%%%%%%%%%%%%%%%%%%%%%%%%%%%%%%%%%%%%%%%%

\begin{figure}[t]
  \includegraphics[width=\linewidth]{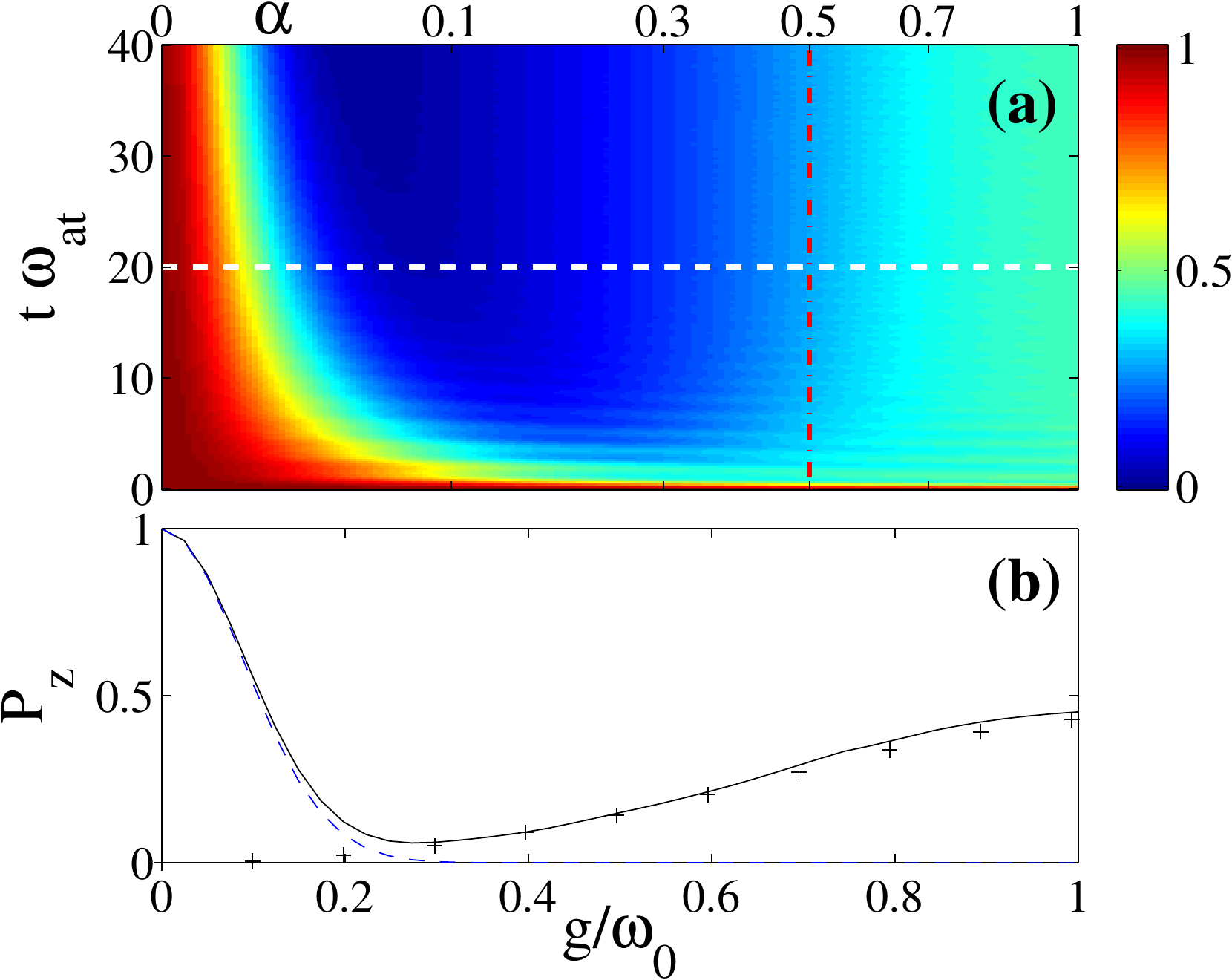}
  \caption{Excitation probability, $P_z = \frac{1}{2}(\langle\sigma_z\rangle+1)$, of an initially excited flux qubit in the open transmission line. (a) Full dynamics and (b) final average excitation at $t\, \omegaat =20$ (black solid), with the master equation prediction (blue dashed) and the ground state value (crosses). $L=121$, $\omegaat=1/3$, $\omega_0=1$. All instances relax to the Bloch vector of the qubit in the ground state. For small values of $g$, however, the relaxation rate decreases as $g^2$ and the excited state probability is a Gaussian.}
  \label{fig:decay}
\end{figure}

We compute the ground state through imaginary time evolution as $\lim_{\tau\to\infty}\ket{\psi(-i\tau)}$, finding that for any coupling $g$ the qubit has some excitation probability $P_z = \frac{1}{2}(\langle \sigma_z\rangle + 1)$~[Fig.\ \ref{fig:decay}b] and the squeezed vacuum polarizes with a nonzero number of photons per site [Fig.\ \ref{fig:ground-state}]. This polarization is evident at $\omegaat=0$, where the ground state becomes a Schroedinger cat with two possible qubit states and a product of displaced coherent states
\begin{equation}
  \ket{\psi_{\omegaat\simeq 0}} \sim \frac{1}{\sqrt{2}} \sum_{s_x=0,1} \ket{s_x} \bigotimes_k \ket{(-1)^{s_x} g u_k/\omega_k},
  \label{polarization}
\end{equation}
that cause a nonzero population of the bosonic modes $\langle n_k\rangle = g^2|u_k|^2/\omega_k^2$.
When we use this ansatz in position space, the distortion is most nonlocal for the $O_x$ coupling. In this case the number of photons per local oscillator departs from the vacuum fluctuations all along the line~[Fig.\ \ref{fig:ground-state}] and the line develops a stationary but small current, $I\propto x_{i+1}-x_i$, directed towards the qubit.

We have studied the spontaneous emission from an excited qubit which is suddenly coupled to the line $\ket{\psi(0)} = \ket{s_z=1} \otimes_k \ket{n_k=0}$. We have found that after a sufficiently long time the state of the system consists of a travelling photon far at the edges of the line, plus a region of the line whose local observables equilibrated together with the qubit. This is first seen for the qubit, whose excitation probability $P_z$, relaxes to that of the combined qubit-line ground state, as shown in Fig.~\ref{fig:decay}b. In the weak and strong coupling limits, $\alpha < 0.1$, the relaxation rate is obtained from a master equation, $\gamma = J(\omega)/2 \propto g^2$. For $g\to 0$, the rate slows down and we find the Gaussian in Fig.~\ref{fig:decay}b. For $\alpha > 0.1$, radiative decay has to be corrected with an asymptotic excitation probability. Finally, for $\alpha > 1/2$ the excited state population relaxes even faster to the ground state value, within a timescale $\sim 1/\omegaat$, deviating from the Markovian law.

\begin{figure}[t]
  \includegraphics[width=\linewidth]{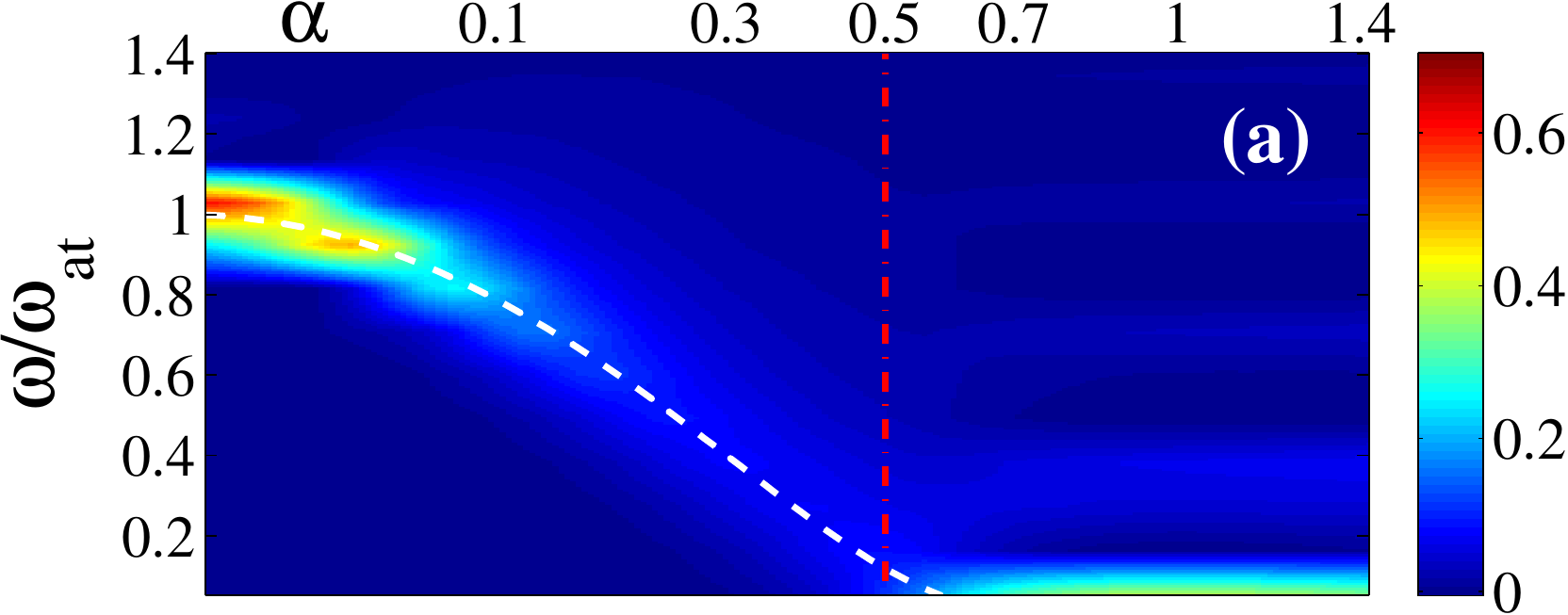}
  \includegraphics[width=\linewidth]{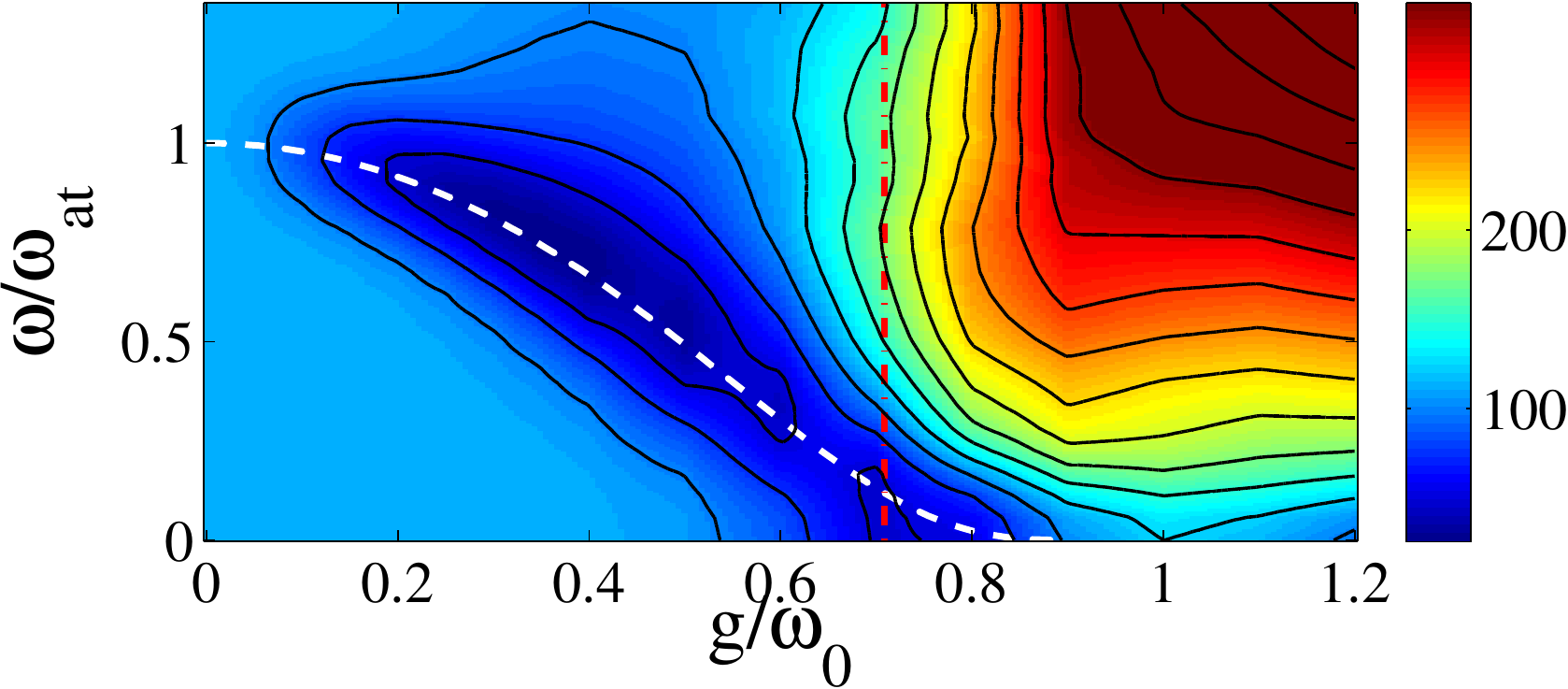}
  \caption{(a) Normalized distribution of photons in frequency space vs. coupling $g/\omega_0$, at $t=45/\omegaat$ after spontaneous emission; RG prediction~(\ref{eq:w-effective}) in dashed and $\alpha=1/2$ line in dash-dot. (b) Transmitted photons for weak coherent wavepacket with average number of photons 1, as a function of the photon frequency, $\omega$, and the qubit-line coupling strength. }
  \label{fig:frequency}
\end{figure}

At the same time that the qubit equilibrates, so do the photons. We have computed the distribution of photons in frequency space, $n_k = \langle a_k^\dagger a_k\rangle$, a long time after the photons are emitted [Fig.\ \ref{fig:frequency}a]. The distribution basically consists on one (or less) extra photons imprinted on top of the state of the line in presence of a qubit. For weak coupling, $\alpha<0.5$, the ground state contains almost no photons and the emitted radiation peaks around
\begin{equation}
  \omegaeff = \omegaat \left( {0.5\omegaat}/{\omega_c}\right)^{\alpha/(1-\alpha)},
  \label{eq:w-effective}
\end{equation}
the resonance estimated in Ref.~\cite{Leggett1987}. For stronger interactions the emitted photon is completely spread in frequency space and $n_k$ is close to the result from Eq.~(\ref{polarization}).

\begin{figure}
  \includegraphics[width=0.9\linewidth]{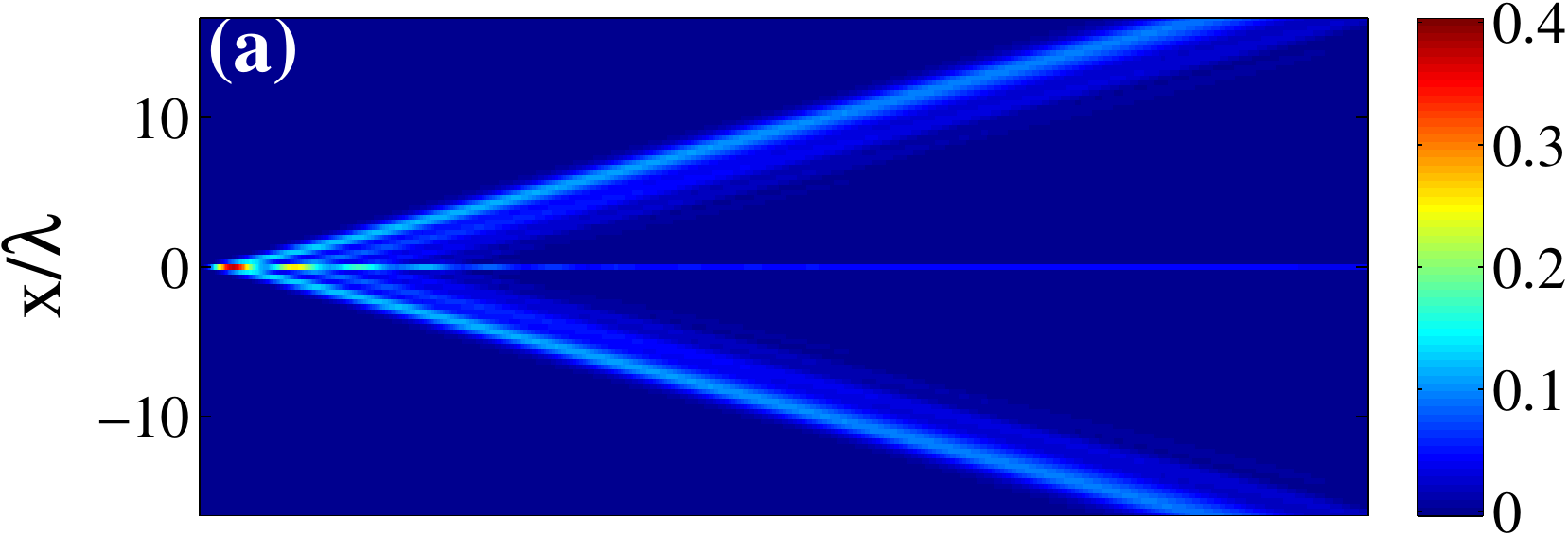}
  \includegraphics[width=0.9\linewidth]{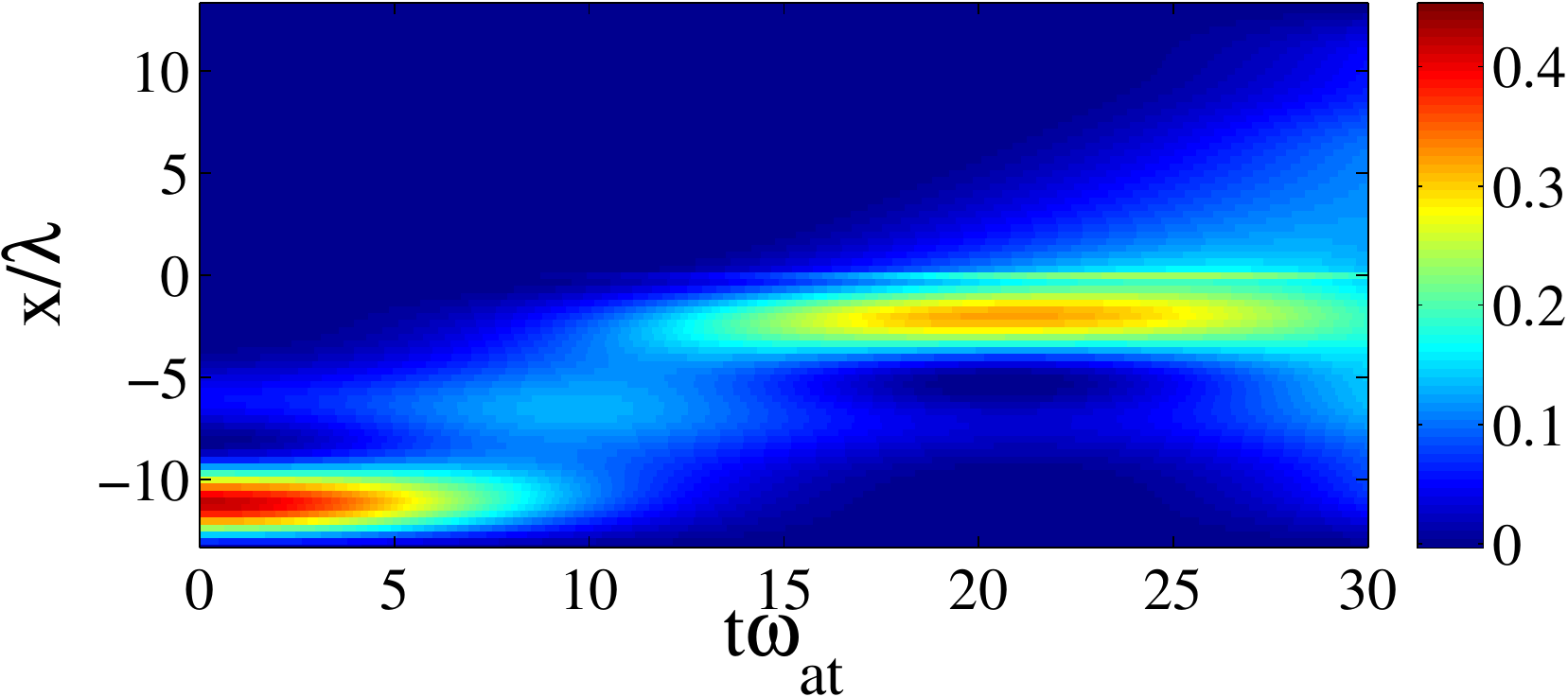}
  \caption{Real space distribution of photons relative to the squeezed vacuum. (a) Spontaneous emission with $g=0.475\omega_0, L=121, \omegaat=1/3$.  (b) Incident and reflected single photon for $g=0.7\omega_0,\,\omega=0.186\omegaat,\, L=121,\, \omegaat=1/3$. }
  \label{fig:position}
\end{figure}

A similar analysis can be done in position space, now studying $\langle \psi(t)| n_i |\psi(t)\rangle- \langle \psi(0)|n_i|\psi(0)\rangle$, the difference between the number of photon per oscillator at times $t$ and zero. This is shown in Fig.\ \ref{fig:position}a for $O_p$ coupling. Note how the travelling photon departs from the qubit leaving the two-level system and its environment in a local state that is close to the ground state. The qubit-line system, though a closed system, seems to provide a bath for its own equilibration, where the bath are the far away extremes of the line. Naturally, this equilibration is incomplete, as finite size effects give rise to revivals due to the photon reflection at the borders, but it is far from obvious that this finite time equilibration works beyond weak system-bath coupling regime.

Remarkably, for all values of the coupling, we still find that the spontaneous emission properties dictate the efficiency of a photon absorption process. In other words, the effective frequency $\omegaeff$ also corresponds to the resonances of the qubit-line system when driven by external photons. We have verified this by studying the interaction of the qubit with a single incident photon. As shown in Fig.~\ref{fig:position}b, a photon of frequency $\omega$ is absorbed and reflected by the qubit after a finite interaction time, a process whose efficiency peaks around $\omega\sim\omegaeff$ [Fig.\ \ref{fig:frequency}a]. Above $\alpha=0.5$, the collision is too broad in time and space, preventing the study of scattering coefficients.

Adding a symmetry breaking perturbation, $\varepsilon\sigma^x/2$, to (\ref{H-position}) allows us to distinguish different thermalization regimes. The latter can be characterized by the susceptibility in the stationary state, $\chi_x = \partial P_x(t\to\infty)/\partial\varepsilon$, with $P_x$ the probability to stay in $| 1 \rangle$ after spontaneous emission. If $\alpha < 1/2$, the steady-state susceptibility, $\chi_x$, matches the ground-state susceptibility of the SB model \cite{LeHur2012}. This result agrees with the fact that in the perturbative regime a Markovian process cools the qubit to the bath temperature. The range $1/2 < \alpha < 1$ corresponds to the antiferromagnetic Kondo phase \cite{Leggett1987}, where the Markovian picture breaks down. Here we find that $\chi_x$ departs from the ground-state value. Finally, in the range $\alpha > 1$, the systems is in the localization phase, such that the qubit dynamics gets frozen and $\chi_x$ vanishes.

\begin{figure}[t]
  \centering
  \includegraphics[width=0.9\linewidth]{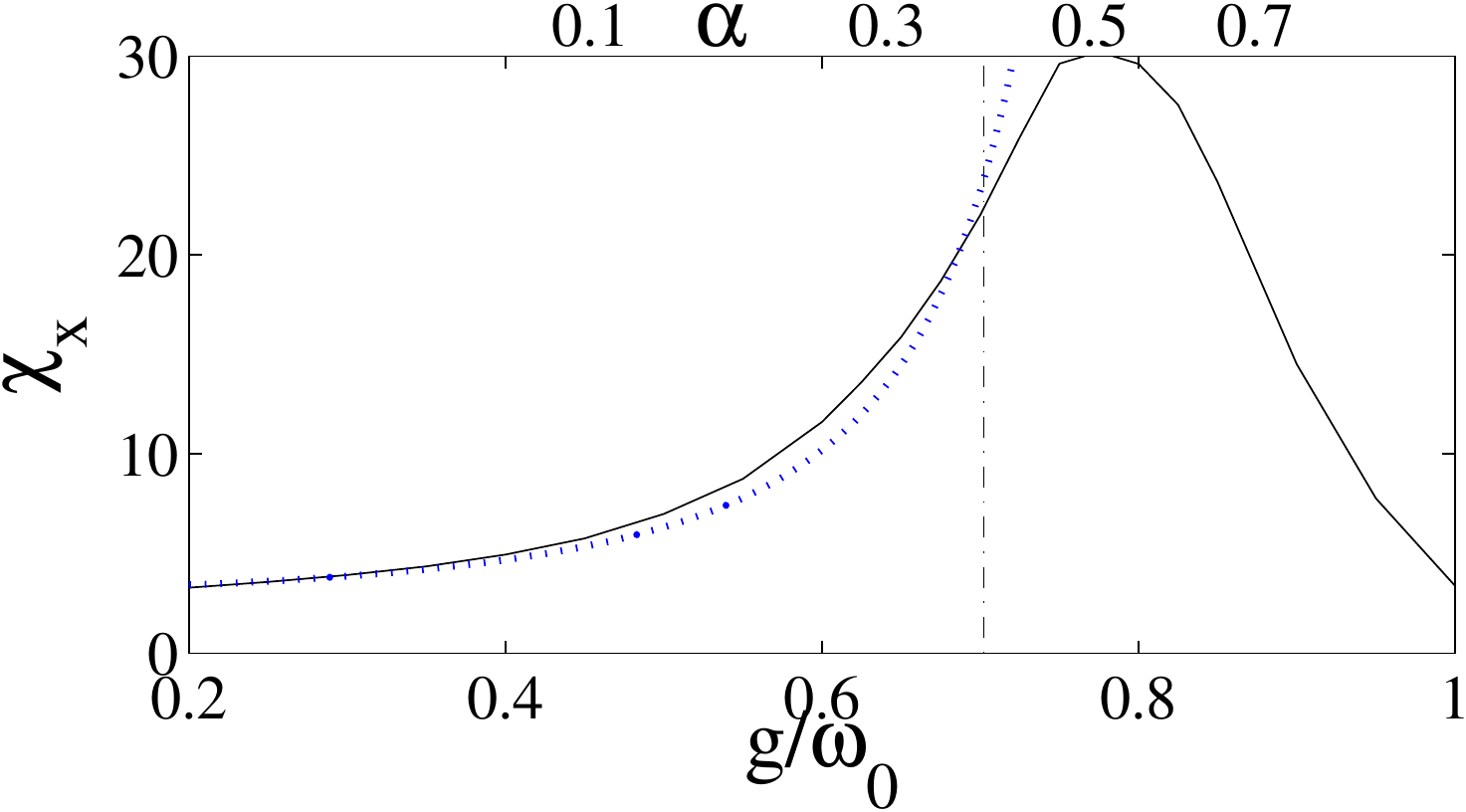}
  \caption{Susceptibility of the qubit along the X direction (solid) when spontaneously emitting a photon while subject to a perturbation $\varepsilon\sigma_x/2$, with a fit to $a/\omegaeff$ (dashed)}
  \label{fig:susceptibility}
\end{figure}

%%%%%%%%%%%%%%%%%%%%%%%%%%%%%%%%%%%%%%%%%%%%%%%%%%%%%%%%%%%%%%%%%%%%%%

\begin{figure}[t]
  \centering
  \includegraphics[width=0.75\linewidth]{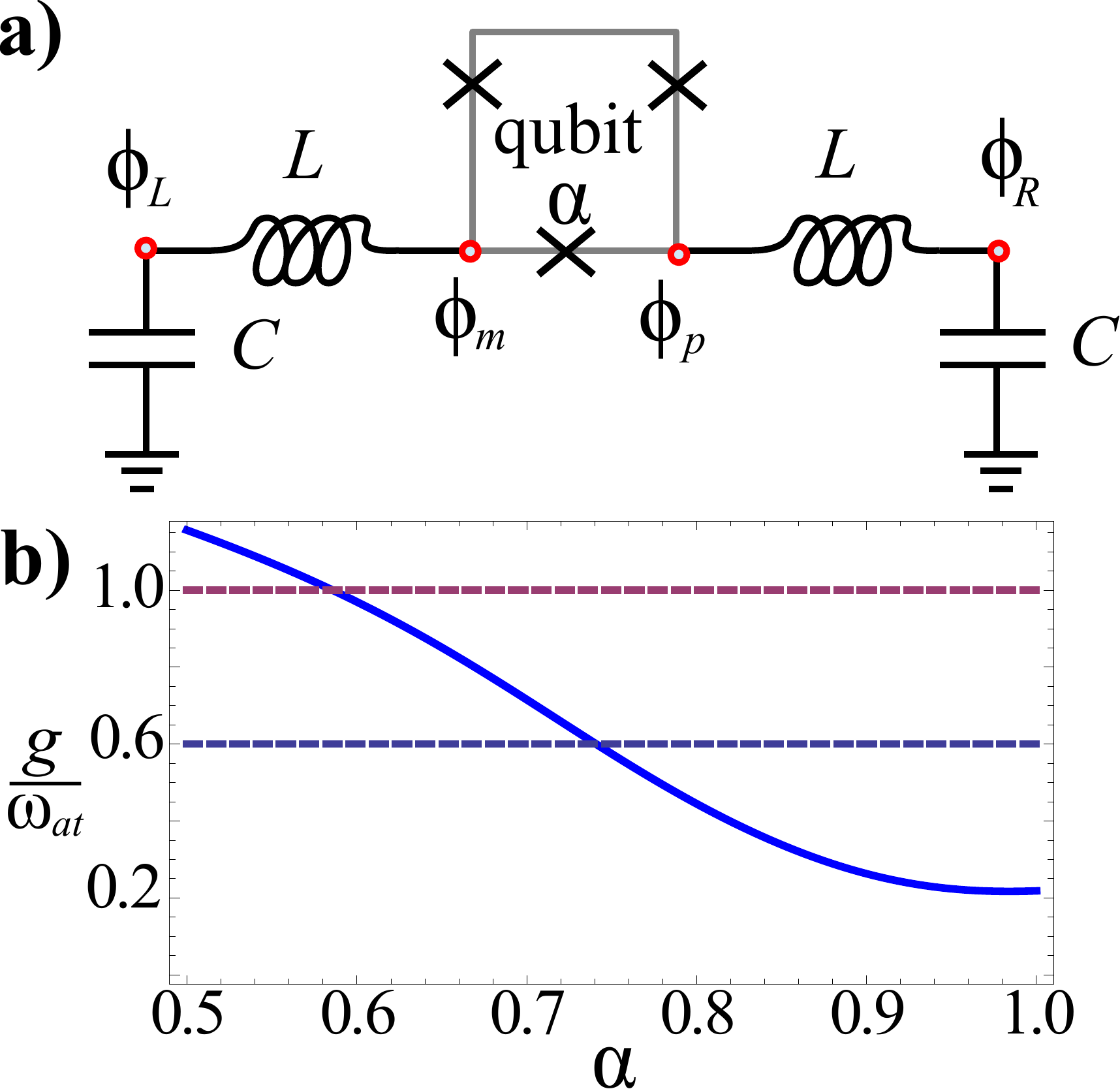}
  \caption{(a) Lumped element circuit for a qubit ultrastrongly coupled to a microwave resonator by inserting it in the transmission line. (b) Effective coupling of a 3-Josephson junction qubit in a line, as a function of the small junction size, $\alpha$. The USC range of $\alpha\sim 0.6-0.7$ is marked in dashed lines.}
  \label{fig:setup}
\end{figure}

We will now discuss how to physically implement the previous model and simulations using a JJ-based qubit in an open transmission line. In previous works~\cite{Bourassa2009}, it was shown that a qubit can be ultrastrongly coupled to a microwave resonator through a derivation that is perturbative in the line-qubit interaction. Instead we now study the qubit as one more element in the discretized transmission line and show that the coupling strength can be obtained non-perturbatively by studying the transmission line and the qubit separately.

Following the setup in Fig.~\ref{fig:setup}a we regard the qubit as a black-box element coupled to an interrupted resonator. We express the full Lagrangian $\mathcal{L} = \mathcal{L}_{LC} + \mathcal{L}_{int} + \mathcal{L}_{qb}'$  in terms of oscillator $\phi_{\pm} := \frac{1}{2}(\phi_R \pm \phi_L)$, and qubit variables $\varphi_{\pm} := \phi_p \pm \phi_m$, obtaining
\begin{eqnarray}
  \mathcal{L}_{LC} &=& \frac{C}{2} \dot\phi_+^2 + \frac{C}{2}\dot\phi_-^2
  - \frac{1}{2L}\phi_+^2 + \frac{1}{2L}\phi_-^2,\\
  \mathcal{L}_{qb}'  &=& \frac{1}{8L}(\varphi_+^2+\varphi_-^2) + \mathcal{L}_{qb}\nonumber\\
  \mathcal{L}_{int} &=& \frac{1}{2L}(\varphi_-\phi_- - \varphi_+\phi_+).\nonumber
\end{eqnarray}
Note the renormalizing effect of the coupling on the qubit, and the simple form of the interaction. In practical examples $\varphi_+$ will not form part of the qubit, but will lock to the oscillator degrees of freedom. We will be thus left with only one operator $\varphi_-$ that couples only to the antisymmetric mode as in Ref.~\cite{Bourassa2009}. Estimating the coupling amounts to computing the matrix elements of $\{\varphi_-,\phi_-\}$ in the resonator and qubit basis.  For this we approximate $\varphi_-$ as a combination of exponentials up to 7-th order, $\varphi_- \simeq \frac{3}{2}\sin(\varphi_-) - \frac{3}{10} \sin(2\varphi_-) + \frac{1}{30}\sin(3\varphi_-)$, and work with them in the number-phase basis.

We have studied the photon-qubit coupling using this model together with a 3-JJ flux qubit and a typical transmission line. As shown in Fig.\ \ref{fig:setup}b, the flux qubit can reach the ultrastrong coupling strength for a moderate value of the small qubit size. These results also show that the coupling strength can be changed using tunable gap qubits that change the effective value of the middle junction~\cite{Paauw2009,Friedman2000,Mooij1999}. This would allow moving in and out of the ultrastrong coupling regime, with the aim of preparing the qubit in excited states and performing the rapid quenches that are needed to do the spontaneous emission experiments. Finally, the same method confirms that the 4-JJ qubit does not achieve a better coupling, and that even an ordinary transmission line can be used for this purpose: there is no substantial need for a constriction, except a fabrication convenience for the embedded qubit.

%%%%%%%%%%%%%%%%%%%%%%%%%%%%%%%%%%%%%%%%%%%%%%%%%%%%%%%%%%%%%%%%%%%%%%

Summing up, in this work we have shown that a flux qubit coupled to an open transmission line implements a quantum simulation of the SB model in the Ohmic regime, with a sufficient range of couplings to cover all regimes: weak, strong, ultrastrong and localization phase. We have developed a numerical method to simulate the physics of such a qubit in the transmission line, including ground state properties, relaxation of the qubit-line system through spontaneous emission, dynamical susceptibility properties and absorption properties. While the numerical methods provide sufficient quantitative evidence and are consistent with some earlier theoretical predictions, an experiment with superconducting qubits would be the only means to provide a definitive confirmation to some of the predictions shown in this work.  The methods put forward in this manuscript will allow the study of more complicated systems, such as the correlation functions\ \cite{Hoi2012} and nonlinear scattering phases\ \cite{Hoi2011} of travelling photons interacting with one or more superconducting qubits, or the effective interactions and entanglement dynamics of qubit ensembles in open transmission lines\ \cite{Gonzalez-Tudela2011}.

\begin{acknowledgements}
The authors acknowledge support from the European project PROMISCE, Spanish MINECO projects FIS2011-25167 and FIS2012-33022, CAM research consortium QUITEMAD (S2009-ESP-1594), and Spanish Ramon y Cajal program
\end{acknowledgements}

%\bibliographystyle{apsrev4-1}
%\bibliography{ultrastrong}

\begin{thebibliography}{27}%
\makeatletter
\providecommand \@ifxundefined [1]{%
 \@ifx{#1\undefined}
}%
\providecommand \@ifnum [1]{%
 \ifnum #1\expandafter \@firstoftwo
 \else \expandafter \@secondoftwo
 \fi
}%
\providecommand \@ifx [1]{%
 \ifx #1\expandafter \@firstoftwo
 \else \expandafter \@secondoftwo
 \fi
}%
\providecommand \natexlab [1]{#1}%
\providecommand \enquote  [1]{``#1''}%
\providecommand \bibnamefont  [1]{#1}%
\providecommand \bibfnamefont [1]{#1}%
\providecommand \citenamefont [1]{#1}%
\providecommand \href@noop [0]{\@secondoftwo}%
\providecommand \href [0]{\begingroup \@sanitize@url \@href}%
\providecommand \@href[1]{\@@startlink{#1}\@@href}%
\providecommand \@@href[1]{\endgroup#1\@@endlink}%
\providecommand \@sanitize@url [0]{\catcode `\\12\catcode `\$12\catcode
  `\&12\catcode `\#12\catcode `\^12\catcode `\_12\catcode `\%12\relax}%
\providecommand \@@startlink[1]{}%
\providecommand \@@endlink[0]{}%
\providecommand \url  [0]{\begingroup\@sanitize@url \@url }%
\providecommand \@url [1]{\endgroup\@href {#1}{\urlprefix }}%
\providecommand \urlprefix  [0]{URL }%
\providecommand \Eprint [0]{\href }%
\providecommand \doibase [0]{http://dx.doi.org/}%
\providecommand \selectlanguage [0]{\@gobble}%
\providecommand \bibinfo  [0]{\@secondoftwo}%
\providecommand \bibfield  [0]{\@secondoftwo}%
\providecommand \translation [1]{[#1]}%
\providecommand \BibitemOpen [0]{}%
\providecommand \bibitemStop [0]{}%
\providecommand \bibitemNoStop [0]{.\EOS\space}%
\providecommand \EOS [0]{\spacefactor3000\relax}%
\providecommand \BibitemShut  [1]{\csname bibitem#1\endcsname}%
\let\auto@bib@innerbib\@empty
%</preamble>
\bibitem [{\citenamefont {Niemczyk}\ \emph {et~al.}(2010)\citenamefont
  {Niemczyk}, \citenamefont {Deppe}, \citenamefont {Huebl}, \citenamefont
  {Menzel}, \citenamefont {Hocke}, \citenamefont {Schwarz}, \citenamefont
  {Garcia-Ripoll}, \citenamefont {Zueco}, \citenamefont {H\"{u}mmer},
  \citenamefont {Solano}, \citenamefont {Marx},\ and\ \citenamefont
  {Gross}}]{Niemczyk2010}%
  \BibitemOpen
  \bibfield  {author} {\bibinfo {author} {\bibfnamefont {T.}~\bibnamefont
  {Niemczyk}}, \bibinfo {author} {\bibfnamefont {F.}~\bibnamefont {Deppe}},
  \bibinfo {author} {\bibfnamefont {H.}~\bibnamefont {Huebl}}, \bibinfo
  {author} {\bibfnamefont {E.~P.}\ \bibnamefont {Menzel}}, \bibinfo {author}
  {\bibfnamefont {F.}~\bibnamefont {Hocke}}, \bibinfo {author} {\bibfnamefont
  {M.~J.}\ \bibnamefont {Schwarz}}, \bibinfo {author} {\bibfnamefont {J.~J.}\
  \bibnamefont {Garcia-Ripoll}}, \bibinfo {author} {\bibfnamefont
  {D.}~\bibnamefont {Zueco}}, \bibinfo {author} {\bibfnamefont
  {T.}~\bibnamefont {H\"{u}mmer}}, \bibinfo {author} {\bibfnamefont
  {E.}~\bibnamefont {Solano}}, \bibinfo {author} {\bibfnamefont
  {A.}~\bibnamefont {Marx}}, \ and\ \bibinfo {author} {\bibfnamefont
  {R.}~\bibnamefont {Gross}},\ }\href {\doibase 10.1038/nphys1730} {\bibfield
  {journal} {\bibinfo  {journal} {Nature Physics}\ }\textbf {\bibinfo {volume}
  {6}},\ \bibinfo {pages} {772} (\bibinfo {year} {2010})}\BibitemShut {NoStop}%
\bibitem [{\citenamefont {Forn-D\'{\i}az}\ \emph {et~al.}(2010)\citenamefont
  {Forn-D\'{\i}az}, \citenamefont {Lisenfeld}, \citenamefont {Marcos},
  \citenamefont {Garc\'{\i}a-Ripoll}, \citenamefont {Solano}, \citenamefont
  {Harmans},\ and\ \citenamefont {Mooij}}]{Forn-Diaz2010}%
  \BibitemOpen
  \bibfield  {author} {\bibinfo {author} {\bibfnamefont {P.}~\bibnamefont
  {Forn-D\'{\i}az}}, \bibinfo {author} {\bibfnamefont {J.}~\bibnamefont
  {Lisenfeld}}, \bibinfo {author} {\bibfnamefont {D.}~\bibnamefont {Marcos}},
  \bibinfo {author} {\bibfnamefont {J.}~\bibnamefont {Garc\'{\i}a-Ripoll}},
  \bibinfo {author} {\bibfnamefont {E.}~\bibnamefont {Solano}}, \bibinfo
  {author} {\bibfnamefont {C.}~\bibnamefont {Harmans}}, \ and\ \bibinfo
  {author} {\bibfnamefont {J.}~\bibnamefont {Mooij}},\ }\href {\doibase
  10.1103/PhysRevLett.105.237001} {\bibfield  {journal} {\bibinfo  {journal}
  {Physical Review Letters}\ }\textbf {\bibinfo {volume} {105}} (\bibinfo
  {year} {2010}),\ 10.1103/PhysRevLett.105.237001}\BibitemShut {NoStop}%
\bibitem [{\citenamefont {G\"{u}nter}\ \emph {et~al.}(2009)\citenamefont
  {G\"{u}nter}, \citenamefont {Anappara}, \citenamefont {Hees}, \citenamefont
  {Sell}, \citenamefont {Biasiol}, \citenamefont {Sorba}, \citenamefont {{De
  Liberato}}, \citenamefont {Ciuti}, \citenamefont {Tredicucci}, \citenamefont
  {Leitenstorfer},\ and\ \citenamefont {Huber}}]{Gunter2009}%
  \BibitemOpen
  \bibfield  {author} {\bibinfo {author} {\bibfnamefont {G.}~\bibnamefont
  {G\"{u}nter}}, \bibinfo {author} {\bibfnamefont {A.~A.}\ \bibnamefont
  {Anappara}}, \bibinfo {author} {\bibfnamefont {J.}~\bibnamefont {Hees}},
  \bibinfo {author} {\bibfnamefont {A.}~\bibnamefont {Sell}}, \bibinfo {author}
  {\bibfnamefont {G.}~\bibnamefont {Biasiol}}, \bibinfo {author} {\bibfnamefont
  {L.}~\bibnamefont {Sorba}}, \bibinfo {author} {\bibfnamefont
  {S.}~\bibnamefont {{De Liberato}}}, \bibinfo {author} {\bibfnamefont
  {C.}~\bibnamefont {Ciuti}}, \bibinfo {author} {\bibfnamefont
  {A.}~\bibnamefont {Tredicucci}}, \bibinfo {author} {\bibfnamefont
  {A.}~\bibnamefont {Leitenstorfer}}, \ and\ \bibinfo {author} {\bibfnamefont
  {R.}~\bibnamefont {Huber}},\ }\href {\doibase 10.1038/nature07838} {\bibfield
   {journal} {\bibinfo  {journal} {Nature}\ }\textbf {\bibinfo {volume}
  {458}},\ \bibinfo {pages} {178} (\bibinfo {year} {2009})}\BibitemShut
  {NoStop}%
\bibitem [{\citenamefont {Anappara}\ \emph {et~al.}(2009)\citenamefont
  {Anappara}, \citenamefont {{De Liberato}}, \citenamefont {Tredicucci},
  \citenamefont {Ciuti}, \citenamefont {Biasiol}, \citenamefont {Sorba},\ and\
  \citenamefont {Beltram}}]{Anappara2009}%
  \BibitemOpen
  \bibfield  {author} {\bibinfo {author} {\bibfnamefont {A.~A.}\ \bibnamefont
  {Anappara}}, \bibinfo {author} {\bibfnamefont {S.}~\bibnamefont {{De
  Liberato}}}, \bibinfo {author} {\bibfnamefont {A.}~\bibnamefont
  {Tredicucci}}, \bibinfo {author} {\bibfnamefont {C.}~\bibnamefont {Ciuti}},
  \bibinfo {author} {\bibfnamefont {G.}~\bibnamefont {Biasiol}}, \bibinfo
  {author} {\bibfnamefont {L.}~\bibnamefont {Sorba}}, \ and\ \bibinfo {author}
  {\bibfnamefont {F.}~\bibnamefont {Beltram}},\ }\href {\doibase
  10.1103/PhysRevB.79.201303} {\bibfield  {journal} {\bibinfo  {journal}
  {Physical Review B}\ }\textbf {\bibinfo {volume} {79}},\ \bibinfo {pages}
  {201303} (\bibinfo {year} {2009})}\BibitemShut {NoStop}%
\bibitem [{\citenamefont {Geiser}\ \emph {et~al.}(2012)\citenamefont {Geiser},
  \citenamefont {Castellano}, \citenamefont {Scalari}, \citenamefont {Beck},
  \citenamefont {Nevou},\ and\ \citenamefont {Faist}}]{Geiser2012}%
  \BibitemOpen
  \bibfield  {author} {\bibinfo {author} {\bibfnamefont {M.}~\bibnamefont
  {Geiser}}, \bibinfo {author} {\bibfnamefont {F.}~\bibnamefont {Castellano}},
  \bibinfo {author} {\bibfnamefont {G.}~\bibnamefont {Scalari}}, \bibinfo
  {author} {\bibfnamefont {M.}~\bibnamefont {Beck}}, \bibinfo {author}
  {\bibfnamefont {L.}~\bibnamefont {Nevou}}, \ and\ \bibinfo {author}
  {\bibfnamefont {J.}~\bibnamefont {Faist}},\ }\href {\doibase
  10.1103/PhysRevLett.108.106402} {\bibfield  {journal} {\bibinfo  {journal}
  {Physical Review Letters}\ }\textbf {\bibinfo {volume} {108}},\ \bibinfo
  {pages} {106402} (\bibinfo {year} {2012})}\BibitemShut {NoStop}%
\bibitem [{\citenamefont {Leggett}\ \emph {et~al.}(1987)\citenamefont
  {Leggett}, \citenamefont {Chakravarty}, \citenamefont {Dorsey}, \citenamefont
  {Fisher}, \citenamefont {Garg},\ and\ \citenamefont {Zwerger}}]{Leggett1987}%
  \BibitemOpen
  \bibfield  {author} {\bibinfo {author} {\bibfnamefont {A.}~\bibnamefont
  {Leggett}}, \bibinfo {author} {\bibfnamefont {S.}~\bibnamefont
  {Chakravarty}}, \bibinfo {author} {\bibfnamefont {A.}~\bibnamefont {Dorsey}},
  \bibinfo {author} {\bibfnamefont {M.}~\bibnamefont {Fisher}}, \bibinfo
  {author} {\bibfnamefont {A.}~\bibnamefont {Garg}}, \ and\ \bibinfo {author}
  {\bibfnamefont {W.}~\bibnamefont {Zwerger}},\ }\href {\doibase
  10.1103/RevModPhys.59.1} {\bibfield  {journal} {\bibinfo  {journal} {Reviews
  of Modern Physics}\ }\textbf {\bibinfo {volume} {59}},\ \bibinfo {pages} {1}
  (\bibinfo {year} {1987})}\BibitemShut {NoStop}%
\bibitem [{\citenamefont {{Le Hur}}(2012)}]{LeHur2012}%
  \BibitemOpen
  \bibfield  {author} {\bibinfo {author} {\bibfnamefont {K.}~\bibnamefont {{Le
  Hur}}},\ }\href {\doibase 10.1103/PhysRevB.85.140506} {\bibfield  {journal}
  {\bibinfo  {journal} {Physical Review B}\ }\textbf {\bibinfo {volume} {85}},\
  \bibinfo {pages} {140506} (\bibinfo {year} {2012})}\BibitemShut {NoStop}%
\bibitem [{\citenamefont {Astafiev}\ \emph {et~al.}(2010)\citenamefont
  {Astafiev}, \citenamefont {Zagoskin}, \citenamefont {Abdumalikov},
  \citenamefont {Pashkin}, \citenamefont {Yamamoto}, \citenamefont {Inomata},
  \citenamefont {Nakamura},\ and\ \citenamefont {Tsai}}]{Astafiev2010}%
  \BibitemOpen
  \bibfield  {author} {\bibinfo {author} {\bibfnamefont {O.}~\bibnamefont
  {Astafiev}}, \bibinfo {author} {\bibfnamefont {A.~M.}\ \bibnamefont
  {Zagoskin}}, \bibinfo {author} {\bibfnamefont {A.~A.}\ \bibnamefont
  {Abdumalikov}}, \bibinfo {author} {\bibfnamefont {Y.~A.}\ \bibnamefont
  {Pashkin}}, \bibinfo {author} {\bibfnamefont {T.}~\bibnamefont {Yamamoto}},
  \bibinfo {author} {\bibfnamefont {K.}~\bibnamefont {Inomata}}, \bibinfo
  {author} {\bibfnamefont {Y.}~\bibnamefont {Nakamura}}, \ and\ \bibinfo
  {author} {\bibfnamefont {J.~S.}\ \bibnamefont {Tsai}},\ }\href {\doibase
  10.1126/science.1181918} {\bibfield  {journal} {\bibinfo  {journal} {Science
  (New York, N.Y.)}\ }\textbf {\bibinfo {volume} {327}},\ \bibinfo {pages}
  {840} (\bibinfo {year} {2010})}\BibitemShut {NoStop}%
\bibitem [{\citenamefont {Abdumalikov}\ \emph {et~al.}(2010)\citenamefont
  {Abdumalikov}, \citenamefont {Astafiev}, \citenamefont {Zagoskin},
  \citenamefont {Pashkin}, \citenamefont {Nakamura},\ and\ \citenamefont
  {Tsai}}]{Abdumalikov2010}%
  \BibitemOpen
  \bibfield  {author} {\bibinfo {author} {\bibfnamefont {A.~A.}\ \bibnamefont
  {Abdumalikov}}, \bibinfo {author} {\bibfnamefont {O.}~\bibnamefont
  {Astafiev}}, \bibinfo {author} {\bibfnamefont {A.~M.}\ \bibnamefont
  {Zagoskin}}, \bibinfo {author} {\bibfnamefont {Y.~A.}\ \bibnamefont
  {Pashkin}}, \bibinfo {author} {\bibfnamefont {Y.}~\bibnamefont {Nakamura}}, \
  and\ \bibinfo {author} {\bibfnamefont {J.~S.}\ \bibnamefont {Tsai}},\ }\href
  {\doibase 10.1103/PhysRevLett.104.193601} {\bibfield  {journal} {\bibinfo
  {journal} {Physical Review Letters}\ }\textbf {\bibinfo {volume} {104}},\
  \bibinfo {pages} {193601} (\bibinfo {year} {2010})}\BibitemShut {NoStop}%
\bibitem [{\citenamefont {Hoi}\ \emph {et~al.}(2011)\citenamefont {Hoi},
  \citenamefont {Wilson}, \citenamefont {Johansson}, \citenamefont {Palomaki},
  \citenamefont {Peropadre},\ and\ \citenamefont {Delsing}}]{Hoi2011}%
  \BibitemOpen
  \bibfield  {author} {\bibinfo {author} {\bibfnamefont {I.-C.}\ \bibnamefont
  {Hoi}}, \bibinfo {author} {\bibfnamefont {C.~M.}\ \bibnamefont {Wilson}},
  \bibinfo {author} {\bibfnamefont {G.}~\bibnamefont {Johansson}}, \bibinfo
  {author} {\bibfnamefont {T.}~\bibnamefont {Palomaki}}, \bibinfo {author}
  {\bibfnamefont {B.}~\bibnamefont {Peropadre}}, \ and\ \bibinfo {author}
  {\bibfnamefont {P.}~\bibnamefont {Delsing}},\ }\href {\doibase
  10.1103/PhysRevLett.107.073601} {\bibfield  {journal} {\bibinfo  {journal}
  {Physical Review Letters}\ }\textbf {\bibinfo {volume} {107}},\ \bibinfo
  {pages} {073601} (\bibinfo {year} {2011})}\BibitemShut {NoStop}%
\bibitem [{\citenamefont {Hoi}\ \emph {et~al.}(2012)\citenamefont {Hoi},
  \citenamefont {Palomaki}, \citenamefont {Lindkvist}, \citenamefont
  {Johansson}, \citenamefont {Delsing},\ and\ \citenamefont
  {Wilson}}]{Hoi2012}%
  \BibitemOpen
  \bibfield  {author} {\bibinfo {author} {\bibfnamefont {I.-C.}\ \bibnamefont
  {Hoi}}, \bibinfo {author} {\bibfnamefont {T.}~\bibnamefont {Palomaki}},
  \bibinfo {author} {\bibfnamefont {J.}~\bibnamefont {Lindkvist}}, \bibinfo
  {author} {\bibfnamefont {G.}~\bibnamefont {Johansson}}, \bibinfo {author}
  {\bibfnamefont {P.}~\bibnamefont {Delsing}}, \ and\ \bibinfo {author}
  {\bibfnamefont {C.~M.}\ \bibnamefont {Wilson}},\ }\href {\doibase
  10.1103/PhysRevLett.108.263601} {\bibfield  {journal} {\bibinfo  {journal}
  {Physical Review Letters}\ }\textbf {\bibinfo {volume} {108}},\ \bibinfo
  {pages} {263601} (\bibinfo {year} {2012})}\BibitemShut {NoStop}%
\bibitem [{\citenamefont {Weichselbaum}\ \emph {et~al.}(2009)\citenamefont
  {Weichselbaum}, \citenamefont {Verstraete}, \citenamefont {Schollw\"{o}ck},
  \citenamefont {Cirac},\ and\ \citenamefont {von Delft}}]{Weichselbaum2009}%
  \BibitemOpen
  \bibfield  {author} {\bibinfo {author} {\bibfnamefont {A.}~\bibnamefont
  {Weichselbaum}}, \bibinfo {author} {\bibfnamefont {F.}~\bibnamefont
  {Verstraete}}, \bibinfo {author} {\bibfnamefont {U.}~\bibnamefont
  {Schollw\"{o}ck}}, \bibinfo {author} {\bibfnamefont {J.}~\bibnamefont
  {Cirac}}, \ and\ \bibinfo {author} {\bibfnamefont {J.}~\bibnamefont {von
  Delft}},\ }\href {\doibase 10.1103/PhysRevB.80.165117} {\bibfield  {journal}
  {\bibinfo  {journal} {Physical Review B}\ }\textbf {\bibinfo {volume} {80}},\
  \bibinfo {pages} {165117} (\bibinfo {year} {2009})}\BibitemShut {NoStop}%
\bibitem [{\citenamefont {Murg}\ \emph {et~al.}(2008)\citenamefont {Murg},
  \citenamefont {Cirac}, \citenamefont {Pirvu},\ and\ \citenamefont
  {Verstraete}}]{Murg2008}%
  \BibitemOpen
  \bibfield  {author} {\bibinfo {author} {\bibfnamefont {V.}~\bibnamefont
  {Murg}}, \bibinfo {author} {\bibfnamefont {J.~I.}\ \bibnamefont {Cirac}},
  \bibinfo {author} {\bibfnamefont {B.}~\bibnamefont {Pirvu}}, \ and\ \bibinfo
  {author} {\bibfnamefont {F.}~\bibnamefont {Verstraete}},\ }\href
  {http://arxiv.org/abs/0804.3976} {\bibfield  {journal} {\bibinfo  {journal}
  {New Journal of Physics}\ }\textbf {\bibinfo {volume} {12}},\ \bibinfo
  {pages} {1} (\bibinfo {year} {2008})}\BibitemShut {NoStop}%
\bibitem [{\citenamefont {Garc\'{\i}a-Ripoll}(2006)}]{Garcia-Ripoll2006}%
  \BibitemOpen
  \bibfield  {author} {\bibinfo {author} {\bibfnamefont {J.~J.}\ \bibnamefont
  {Garc\'{\i}a-Ripoll}},\ }\href {\doibase 10.1088/1367-2630/8/12/305}
  {\bibfield  {journal} {\bibinfo  {journal} {New Journal of Physics}\ }\textbf
  {\bibinfo {volume} {8}},\ \bibinfo {pages} {305} (\bibinfo {year}
  {2006})}\BibitemShut {NoStop}%
\bibitem [{\citenamefont {Anders}\ and\ \citenamefont
  {Schiller}(2005)}]{Anders2005}%
  \BibitemOpen
  \bibfield  {author} {\bibinfo {author} {\bibfnamefont {F.~B.}\ \bibnamefont
  {Anders}}\ and\ \bibinfo {author} {\bibfnamefont {A.}~\bibnamefont
  {Schiller}},\ }\href {\doibase 10.1103/PhysRevLett.95.196801} {\bibfield
  {journal} {\bibinfo  {journal} {Phys. Rev. Lett.}\ }\textbf {\bibinfo
  {volume} {95}},\ \bibinfo {pages} {196801} (\bibinfo {year}
  {2005})}\BibitemShut {NoStop}%
\bibitem [{\citenamefont {Anders}\ and\ \citenamefont
  {Schiller}(2006)}]{Anders2006}%
  \BibitemOpen
  \bibfield  {author} {\bibinfo {author} {\bibfnamefont {F.~B.}\ \bibnamefont
  {Anders}}\ and\ \bibinfo {author} {\bibfnamefont {A.}~\bibnamefont
  {Schiller}},\ }\href {\doibase 10.1103/PhysRevB.74.245113} {\bibfield
  {journal} {\bibinfo  {journal} {Phys. Rev. B}\ }\textbf {\bibinfo {volume}
  {74}},\ \bibinfo {pages} {245113} (\bibinfo {year} {2006})}\BibitemShut
  {NoStop}%
\bibitem [{\citenamefont {Orth}\ \emph {et~al.}(2010)\citenamefont {Orth},
  \citenamefont {Roosen}, \citenamefont {Hofstetter},\ and\ \citenamefont
  {Le~Hur}}]{Orth2010}%
  \BibitemOpen
  \bibfield  {author} {\bibinfo {author} {\bibfnamefont {P.~P.}\ \bibnamefont
  {Orth}}, \bibinfo {author} {\bibfnamefont {D.}~\bibnamefont {Roosen}},
  \bibinfo {author} {\bibfnamefont {W.}~\bibnamefont {Hofstetter}}, \ and\
  \bibinfo {author} {\bibfnamefont {K.}~\bibnamefont {Le~Hur}},\ }\href@noop {}
  {\bibfield  {journal} {\bibinfo  {journal} {Physical Review B}\ }\textbf
  {\bibinfo {volume} {82}},\ \bibinfo {pages} {144423} (\bibinfo {year}
  {2010})}\BibitemShut {NoStop}%
\bibitem [{\citenamefont {Prior}\ \emph {et~al.}(2010)\citenamefont {Prior},
  \citenamefont {Chin}, \citenamefont {Huelga},\ and\ \citenamefont
  {Plenio}}]{Prior2010}%
  \BibitemOpen
  \bibfield  {author} {\bibinfo {author} {\bibfnamefont {J.}~\bibnamefont
  {Prior}}, \bibinfo {author} {\bibfnamefont {A.~W.}\ \bibnamefont {Chin}},
  \bibinfo {author} {\bibfnamefont {S.~F.}\ \bibnamefont {Huelga}}, \ and\
  \bibinfo {author} {\bibfnamefont {M.~B.}\ \bibnamefont {Plenio}},\
  }\href@noop {} {\bibfield  {journal} {\bibinfo  {journal} {Physical review
  letters}\ }\textbf {\bibinfo {volume} {105}},\ \bibinfo {pages} {050404}
  (\bibinfo {year} {2010})}\BibitemShut {NoStop}%
\bibitem [{\citenamefont {Bourassa}\ \emph {et~al.}(2009)\citenamefont
  {Bourassa}, \citenamefont {Gambetta}, \citenamefont {Abdumalikov},
  \citenamefont {Astafiev}, \citenamefont {Nakamura},\ and\ \citenamefont
  {Blais}}]{Bourassa2009}%
  \BibitemOpen
  \bibfield  {author} {\bibinfo {author} {\bibfnamefont {J.}~\bibnamefont
  {Bourassa}}, \bibinfo {author} {\bibfnamefont {J.}~\bibnamefont {Gambetta}},
  \bibinfo {author} {\bibfnamefont {A.}~\bibnamefont {Abdumalikov}}, \bibinfo
  {author} {\bibfnamefont {O.}~\bibnamefont {Astafiev}}, \bibinfo {author}
  {\bibfnamefont {Y.}~\bibnamefont {Nakamura}}, \ and\ \bibinfo {author}
  {\bibfnamefont {A.}~\bibnamefont {Blais}},\ }\href {\doibase
  10.1103/PhysRevA.80.032109} {\bibfield  {journal} {\bibinfo  {journal}
  {Physical Review A}\ }\textbf {\bibinfo {volume} {80}},\ \bibinfo {pages}
  {032109} (\bibinfo {year} {2009})}\BibitemShut {NoStop}%
\bibitem [{\citenamefont {Denker}(1984)}]{Denker1984}%
  \BibitemOpen
  \bibfield  {author} {\bibinfo {author} {\bibfnamefont {J.~S.}\ \bibnamefont
  {Denker}},\ }\href@noop {} {\  (\bibinfo {year} {1984})}\BibitemShut
  {NoStop}%
\bibitem [{\citenamefont {Devoret}(1995)}]{Devoret1995}%
  \BibitemOpen
  \bibfield  {author} {\bibinfo {author} {\bibfnamefont {M.}~\bibnamefont
  {Devoret}},\ }\href
  {http://www.physique.usherb.ca/tremblay/cours/PHY-731/Quantum\_circuit\_theory-1.pdf}
  {\bibfield  {journal} {\bibinfo  {journal} {Les Houches, Session LXIII}\ ,\
  \bibinfo {pages} {351}} (\bibinfo {year} {1995})}\BibitemShut {NoStop}%
\bibitem [{Note1()}]{Note1}%
  \BibitemOpen
  \bibinfo {note} {All simulations have been done using both couplings, without
  significant differences except where explicitly noted.}\BibitemShut {Stop}%
\bibitem [{Note2()}]{Note2}%
  \BibitemOpen
  \bibinfo {note} {The entanglement generated by the coupling leads to a large
  number of photons $n_i=\protect \genfrac {}{}{}1{1}{2}(x_i^2+p_i^2-1)$ that
  demand a large cut-off to represent the state faithfully, even without a
  qubit.}\BibitemShut {Stop}%
\bibitem [{\citenamefont {Paauw}\ \emph {et~al.}(2009)\citenamefont {Paauw},
  \citenamefont {Fedorov}, \citenamefont {Harmans},\ and\ \citenamefont
  {Mooij}}]{Paauw2009}%
  \BibitemOpen
  \bibfield  {author} {\bibinfo {author} {\bibfnamefont {F.}~\bibnamefont
  {Paauw}}, \bibinfo {author} {\bibfnamefont {A.}~\bibnamefont {Fedorov}},
  \bibinfo {author} {\bibfnamefont {C.}~\bibnamefont {Harmans}}, \ and\
  \bibinfo {author} {\bibfnamefont {J.}~\bibnamefont {Mooij}},\ }\href
  {\doibase 10.1103/PhysRevLett.102.090501} {\bibfield  {journal} {\bibinfo
  {journal} {Physical Review Letters}\ }\textbf {\bibinfo {volume} {102}},\
  \bibinfo {pages} {090501} (\bibinfo {year} {2009})}\BibitemShut {NoStop}%
\bibitem [{\citenamefont {Friedman}\ \emph {et~al.}(2000)\citenamefont
  {Friedman}, \citenamefont {Patel}, \citenamefont {Chen}, \citenamefont
  {Tolpygo},\ and\ \citenamefont {Lukens}}]{Friedman2000}%
  \BibitemOpen
  \bibfield  {author} {\bibinfo {author} {\bibfnamefont {J.}~\bibnamefont
  {Friedman}}, \bibinfo {author} {\bibfnamefont {V.}~\bibnamefont {Patel}},
  \bibinfo {author} {\bibfnamefont {W.}~\bibnamefont {Chen}}, \bibinfo {author}
  {\bibfnamefont {S.}~\bibnamefont {Tolpygo}}, \ and\ \bibinfo {author}
  {\bibfnamefont {J.}~\bibnamefont {Lukens}},\ }\href {\doibase
  10.1038/35017505} {\bibfield  {journal} {\bibinfo  {journal} {Nature}\
  }\textbf {\bibinfo {volume} {406}},\ \bibinfo {pages} {43} (\bibinfo {year}
  {2000})}\BibitemShut {NoStop}%
\bibitem [{\citenamefont {Mooij}\ \emph {et~al.}(1999)\citenamefont {Mooij},
  \citenamefont {Orlando}, \citenamefont {Levitov}, \citenamefont {Tian},
  \citenamefont {van~der Wal},\ and\ \citenamefont {Lloyd}}]{Mooij1999}%
  \BibitemOpen
  \bibfield  {author} {\bibinfo {author} {\bibfnamefont {J.~E.}\ \bibnamefont
  {Mooij}}, \bibinfo {author} {\bibfnamefont {T.~P.}\ \bibnamefont {Orlando}},
  \bibinfo {author} {\bibfnamefont {L.}~\bibnamefont {Levitov}}, \bibinfo
  {author} {\bibfnamefont {L.}~\bibnamefont {Tian}}, \bibinfo {author}
  {\bibfnamefont {C.~H.}\ \bibnamefont {van~der Wal}}, \ and\ \bibinfo {author}
  {\bibfnamefont {S.}~\bibnamefont {Lloyd}},\ }\href {\doibase
  10.1126/science.285.5430.1036} {\bibfield  {journal} {\bibinfo  {journal}
  {Science}\ }\textbf {\bibinfo {volume} {285}},\ \bibinfo {pages} {1036}
  (\bibinfo {year} {1999})}\BibitemShut {NoStop}%
\bibitem [{\citenamefont {Gonzalez-Tudela}\ \emph {et~al.}(2011)\citenamefont
  {Gonzalez-Tudela}, \citenamefont {Martin-Cano}, \citenamefont {Moreno},
  \citenamefont {Martin-Moreno}, \citenamefont {Tejedor},\ and\ \citenamefont
  {Garcia-Vidal}}]{Gonzalez-Tudela2011}%
  \BibitemOpen
  \bibfield  {author} {\bibinfo {author} {\bibfnamefont {A.}~\bibnamefont
  {Gonzalez-Tudela}}, \bibinfo {author} {\bibfnamefont {D.}~\bibnamefont
  {Martin-Cano}}, \bibinfo {author} {\bibfnamefont {E.}~\bibnamefont {Moreno}},
  \bibinfo {author} {\bibfnamefont {L.}~\bibnamefont {Martin-Moreno}}, \bibinfo
  {author} {\bibfnamefont {C.}~\bibnamefont {Tejedor}}, \ and\ \bibinfo
  {author} {\bibfnamefont {F.~J.}\ \bibnamefont {Garcia-Vidal}},\ }\href
  {\doibase 10.1103/PhysRevLett.106.020501} {\bibfield  {journal} {\bibinfo
  {journal} {Physical Review Letters}\ }\textbf {\bibinfo {volume} {106}},\
  \bibinfo {pages} {020501} (\bibinfo {year} {2011})}\BibitemShut {NoStop}%
\end{thebibliography}


\end{document}